\documentclass{article}%
\usepackage{geometry}
\usepackage{amsmath}
\usepackage{cite}
\usepackage{color}
\usepackage{soul}
\usepackage{amsfonts}
\usepackage{amssymb}
\usepackage{booktabs}
\usepackage{array}
\usepackage{graphicx}
\usepackage{float}%
\providecommand{\U}[1]{\protect\rule{.1in}{.1in}}
\newtheorem{theorem}{Theorem}

\newcolumntype{M}{>{\centering\arraybackslash}m{\dimexpr.25\linewidth-2\tabcolsep}}
\begin{document}

\title{Performance measures for single-degree-of-freedom energy harvesters under
stochastic excitation}
\author{Han Kyul Joo and Themistoklis P.\ Sapsis\thanks{Corresponding author:
sapsis@mit.edu, Tel: +1 617-324-7508, Fax: +1 617-253-8689}\\Massachusetts Institute of Technology, 77 Massachusetts Av, Cambridge MA 02139}
\date{\today}
\maketitle

\begin{abstract}
We develop performance criteria for the objective comparison of different
classes of single-degree-of-freedom oscillators under stochastic excitation.
For each family of oscillators, these objective criteria take into account the
maximum possible energy harvested for a given response level, which is a
quantity that is directly connected to the size of the harvesting
configuration. We prove that the derived criteria are invariant with respect
to magnitude or temporal rescaling of the input spectrum and they depend only
on the relative distribution of energy across different harmonics of the
excitation. We then compare three different classes of linear and nonlinear
oscillators and using stochastic analysis methods we illustrate that in all
cases of excitation spectra (monochromatic, broadband, white-noise) the
optimal performance of all designs cannot exceed the performance of the linear
design. Subsequently, we study the robustness of this optimal performance to
small perturbations of the input spectrum and illustrate the advantages of
nonlinear designs relative to linear ones.

\end{abstract}

\section{Introduction}

Energy harvesting is the process of targeted energy transfer from a given
source (e.g. ambient mechanical vibrations, water waves, etc) to specific
dynamical modes with the aim of transforming this energy to useful forms (e.g.
electricity). In general, a source of mechanical energy can be described in
terms of the displacement, velocity or acceleration spectrum. Moreover, in
most cases the existence of the energy harvesting device does not alter the
properties of the energy source i.e. the device is essentially driven by the
energy source in a one-way interaction.

Typical energy sources are usually characterized by non-monochromatic energy
content, i.e. the energy is spread over a finite band of frequencies. This
feature has led to the development of various techniques in order to achieve
efficient energy harvesting. Many of these approaches employ
single-degree-of-freedom oscillators with non-quadratic potentials, i.e. with
a restoring force that is nonlinear see e.g. \cite{Daqaq10, Daqaq11, Green12,
Stephen06, Halvorsen13, Barton10, Green13, harne13, Mendez13, Brennan14, Watt94, McInnes08, Mann09}. In all of these
approaches, a common characteristic is the employment of intensional
nonlinearity in the harvester dynamics with an ultimate scope of increasing
performance and robustness of the device without changing its size, mass or
the amount of its kinetic energy. Even though for linear systems the response
of the harvester can be fully characterized (and therefore optimized) in terms
of the energy-source spectrum (see e.g. \cite{Stephen06, Mendez13linear}),
this is not the case for nonlinear systems which are simultaneously excited by
multiple harmonics - in this case there are no analytical methods to express the
stochastic response in terms of the source spectrum. While in many cases (e.g.
in \cite{Green12, Barton10, harne13, Mann09}) the authors observe clear indications
that the energy harvesting capacity is increased in the presence of
nonlinearity, in numerous other studies (e.g. \cite{Daqaq11, Daqaq10,
Halvorsen13, Green13}) these benefits could not be observed. To this end it is
not obvious if and when a class (i.e. a family) of nonlinear energy harvesters
can perform ``better" relative to another class (of linear or nonlinear
systems) of energy harvesters when these are excited by a given source spectrum.

Here we seek to define objective criteria that will allow us to choose an
optimal and robust energy harvester design for a given energy source spectrum.
An efficient energy harvester (EH) can be informally defined as the
configuration that is able to harvest the largest possible amount of
energy\textit{ for a given size and mass}. This is a particularly challenging
question since the performance of any given design depends strongly on the
chosen system parameters (e.g. damping, stiffness, etc.) and in order to
compare different classes of systems (e.g. linear versus nonlinear) the
developed measures should not depend on the specific system parameters but
rather on the form of the design, its size or mass as well as the energy
source spectrum. Similar challenges {araise} when one tries to quantify the
robustness of a given design to variations of the source spectrum for which it
has been optimized.

To pursue this goal we first develop measures that quantify the performance of
general nonlinear systems from broadband spectra, i.e. simultaneous excitation
from a broad range of harmonics. These criteria demonstrate for each class of
systems the maximum possible power that can be harvested from a fixed energy
source using a given volume. We prove that the developed measures are
invariant to linear transformations of the source spectrum (i.e. rescaling in
time and size of the excitation) and they essentially depend only on its
shape, i.e. the relative distribution of energy among different harmonics. For
the sake of simplicity, we will present our measures for one dimensional
systems although they can be generalized to higher dimensional cases in a
straightforward manner.

Using the derived criteria we examine the relative advantages of different
classes of single-degree-of-freedom (SDOF) harvesters. We examine various
extreme scenarios of source spectra ranging from monochromatic excitations to
white-noise cases (also including the intermediate case of the
Pierson-Moskowitz (PM) spectrum). We prove that there are fundamental
limitations on the maximum possible harvested power that can be achieved
(using SDOF harvesters) and these are independent from the linear or nonlinear
nature of the design. Moreover, we examine the robustness properties of
various SDOF harvester designs when the source characteristics are perturbed
and we illustrate the dynamical regimes where non-linear designs are
preferable compared with the linear harvesters.

\section{Quantification of power harvesting performance under broadband
excitation}

We study the energy harvesting properties of a SDOF oscillator subjected to
random excitation. In the energy harvesting setting, randomness is usually
introduced through the excitation signal which although is characterized by a
given spectrum, i.e. a given amplitude for each harmonic, the relative phase
between harmonics is unknown and to this end is modeled as a uniformly
distributed random variable\textit{. }We consider the following system
consisting of an oscillator lying on a basis whose displacement $h\left(
t\right)  $ is a random function of time with given spectrum. The equation of
motion for this simple system has the form%
\begin{equation}
m\ddot{x}+\lambda\left(  \dot{x}-\dot{h}\right)  +F\left(  x-h\right)  =0,
\label{system1}%
\end{equation}
where $m$ is the mass of the system, $\lambda$ is a dissipation coefficient
expressing only the harvesting of energy (we ignore in this simple setting any
mechanical loses), and $F$ is the spring force that has a given form but free
parameters, i.e. $F\left(  x\right)  =F\left(  x;k_{1},...,k_{n}\right)  .$
One could think of $F$ as a polynomial: $F\left(  x;k_{p}\right)
={\sum\limits_{p=1,...,N}}k_{p}x^{p}$.

We assume that the excitation process is stationary and ergodic having a given
spectrum $S_{hh}\left(  \omega\right)  $ (See Appendix I for definition)$.$ We
also assume that after sufficient time the system converges to a statistical
steady state where the response can be characterized by the power spectrum
$S_{qq}\left(  \omega\right)  .$ For this system the harvested power per unit
mass is given by
\begin{equation}
P_{h}=\frac{\lambda}{m}\overline{\left(  \dot{x}-\dot{h}\right)  ^{2}},
\end{equation}
where the bar denotes ensemble or temporal average in the statistical steady
state regime of the dynamics. For convenience we apply the transformation
$x-h=q$ to obtain the system%
\begin{equation}
\ddot{q}+\hat{\lambda}\dot{q}+\hat{F}\left(  q\right)  =-\ddot{h},
\label{Eq_org_trans}%
\end{equation}
where $\hat{\lambda}=\frac{\lambda}{m}$ and $\hat{F}=\frac{F}{m}.$

Through this formulation we note that the mass can be regarded as a parameter
that does not need to be taken into account in the optimization procedure.
This is because for any optimal set of parameters $\hat{\lambda}$ and $\hat
{F}$, the energy harvested will increase linearly with the mass of the
oscillator employed (given that $\hat{\lambda}$ and $\hat{F}$ remain constant).

\subsection{Absolute and normalized harvested power $P_{h}$}

In the present work, we are interested to compare the maximum possible
performance between different classes of oscillators and to this end we ignore
mechanical losses and assume that the damping coefficient $\hat{\lambda}$
describes entirely the energy harvested. In terms of the spectral properties
of the response, the absolute harvested power $P_{h}$ can then be expressed as%
\begin{equation}
P_{h}=\hat{\lambda}\overline{\dot{q}^{2}}=\hat{\lambda}%
{\displaystyle\int\limits_{-\infty}^{\infty}}
\omega^{2}S_{qq}\left(  \omega\right)  d\omega.
\end{equation}
This quantifies the amount of energy harvested per unit mass.

\subsection{Size of the energy harvester $\mathcal{B}$}

An objective comparison between two harvesters should involve not only the
same mass but also the same size. We chose to quantify the characteristic size
of the harvesting device using the mean square displacement of the center of
mass of the system. For the SDOF setting, this is simply the typical deviation
of the stochastic process $q\left(  t\right)  $ given by
\begin{equation}
d=\sqrt{\overline{q^{2}}}=\sqrt{%
{\displaystyle\int\limits_{-\infty}^{\infty}}
S_{qq}\left(  \omega\right)  d\omega}.
\end{equation}
Our goal is to quantify the maximum performance of a harvesting configuration
for a given typical size $d$ and for a given form of input spectrum. To
achieve invariance with respect to the source-spectrum magnitude, we will use
the non-dimensional ratio
\begin{equation}
\mathcal{B}=\frac{\overline{q^{2}}}{\overline{h^{2}}},
\end{equation}
which is the square of the relative magnitude of the device compared with the
typical size of the excitation motion $\sqrt{\overline{h^{2}}}$. The above
quantity also expresses the amount of energy that the device carries relative
to the energy of the excitation and to this end we will refer to it as the
response level of the harvester. It will be used to parametrize the
performance measures developed in the next section with respect to the typical
size of the device.

\subsection{Harvested power density $\rho_{e}$}

For each response level $\mathcal{B}$, we define the\textit{ harvested power
density} $\rho_{e}$ as the maximum possible harvested power
$\underset{\left\{  \hat{\lambda},\hat{k}_{i}\text{ }|\text{ }\mathcal{B}%
\right\}  }{\max}P_{h}$ (for a given excitation spectrum and under the
{constraint} of a given response level $\mathcal{B}$) suitably normalized with
respect to the response size $\overline{q^{2}}$ and the mean frequency of the
input spectrum
\begin{equation}
\rho_{e}(\mathcal{B})=\frac{\underset{\left\{  \hat{\lambda}, \hat{k}_{i}\text{
}|\text{ }\mathcal{B}\right\}  }{\max}P_{h}}{\omega_{h}^{3}\overline{q^{2}}%
}=\frac{\underset{\left\{  \hat{\lambda}, \hat{k}_{i}\text{ }|\text{
}\mathcal{B}\right\}  }{\max}\left(  \hat{\lambda}\overline{\dot{q}^{2}%
}\right)  }{\omega_{h}^{3}\overline{q^{2}}}%
\end{equation}
where the mean frequency of the input spectrum is defined as%
\begin{equation}
\omega_{h}=\frac{1}{\overline{h^{2}}}{\int\limits_{0}^{\infty}}\omega
S_{hh}\left(  \omega\right)  d\omega.
\end{equation}
This measure should be viewed as a function of the response level of the
device $\mathcal{B}$. As we show below it satisfies an invariance property
under linear transformations of the excitation spectrum, i.e. rescaling of the
spectrum in time and magnitude (Figure 1). More specifically we have the
following theorem.%
\begin{figure}[H]%
\centering
\includegraphics[
height=2.802in,
width=3.6037in
]%
{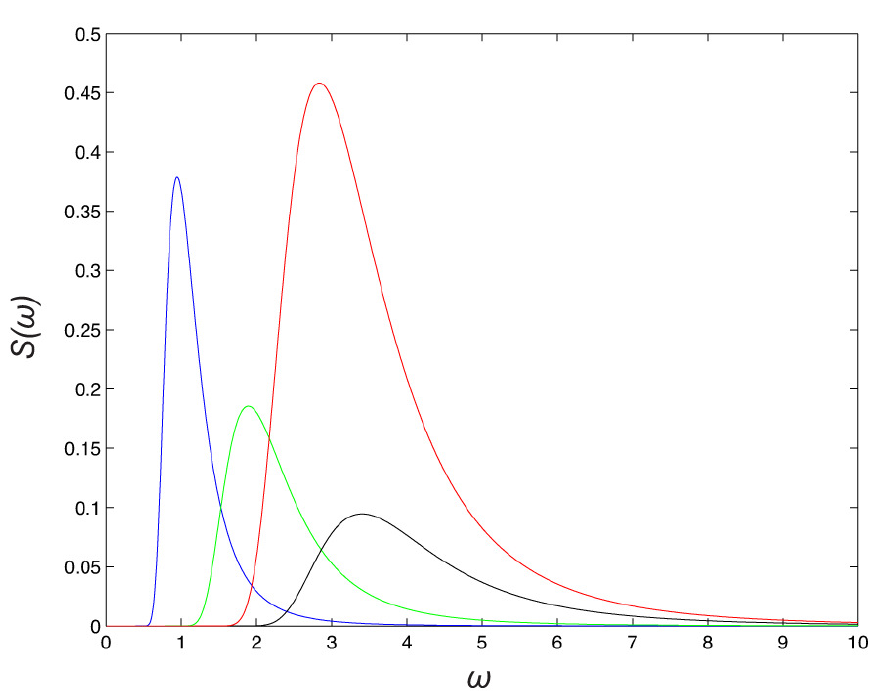}%
\caption{Various spectral curves obtained by magnitude and temporal rescaling
of the Pierson-Moskowitz spectrum. Amplification and stretching of the input
spectrum will leave the effective damping and the harvested power density
invariant.}%
\label{fig_spectra}%
\end{figure}

\begin{theorem}
\textit{The harvested power density }$\rho_{e}$ \textit{is invariant with
respect to linear transformations of the input energy spectrum }$S_{hh}\left(
\omega\right)  $ \textit{(uniform amplification and stretching). In
particular, under the modified excitation }$g\left(  t\right)  =a\sqrt
{b}h\left(  bt\right)  $ or equivalently the input spectrum $S_{gg}\left(
\omega\right)  =a^{2}S_{hh}\left(  \frac{\omega}{b}\right)  $, where\textit{
}$a>0$ and $b>0$ , the curve $\rho_{e}\left(  \mathcal{B}\right)  $ remain invariant.
\end{theorem}

\noindent
\textbf{Proof.}
Let $\hat{\lambda}_{0}$ and $\hat{k}_{i,0}$ be the optimal parameters for
which the quantity $P_{h}$ attains its maximum value for the input spectrum
$S_{hh}\left(  \omega\right)  $ under the constraint of a given response level
$\mathcal{B}_{0}=\frac{\overline{q^{2}}}{\overline{h^{2}}}.$ For convenience,
we will use the notation $\hat{F}_{0}\left(  q\right)  =\hat{F}\left(
q;\hat{k}_{1,0},...,\hat{k}_{n,0}\right)  .$ For these optimal parameters we
will also have the optimum response $q_{0}\left(  t\right)  $ that satisfies
the equation%
\begin{equation}
\ddot{q}_{0}+\hat{\lambda}_{0}\dot{q}_{0}+\hat{F}_{0}\left(  q_{0}\right)
=-\ddot{h}. \label{eq_orgn}%
\end{equation}
We will prove that under the rescaled spectrum $S_{gg}\left(  \omega\right)
=a^{2}S_{hh}\left(  \frac{\omega}{b}\right)  $ the harvested power density
curve $\rho_{e}\left(  \mathcal{B}\right)  $ remains invariant.\ By direct
computation, it can be verified that the modified spectrum $S_{gg}\left(
\omega\right)  $ corresponds to an excitation of the form%
\begin{equation}
g\left(  t\right)  =a\sqrt{b}h\left(  bt\right) .
\end{equation}
Moreover, by direct calculation we can verify that%
\begin{equation}
\overline{g^{2}}=a^{2}b\overline{h^{2}}\text{ \ \ \ and \ \ \ }\omega
_{g}=b\omega_{h}.
\end{equation}
We pick a response level $\mathcal{B}_{0}$ for the system excited by $h\left(
t\right)  \ $and we will prove that $\rho_{e,g}\left(  \mathcal{B}_{0}\right)
=\rho_{e,h}\left(  \mathcal{B}_{0}\right)  .$ Under the new excitation the
system equation will be
\begin{equation}
\ddot{q}+\hat{\lambda}\dot{q}+\hat{F}\left(  q\right)  =-a\sqrt{b}\frac
{d^{2}h\left(  bt\right)  }{dt^{2}}. \label{eq_one1}%
\end{equation}
We apply the temporal transformation $bt=\tau.$ In the new timescale, we will
have (differentiation is now denoted with $^{\prime}$)%
\begin{equation}
b^{2}q^{\prime\prime}+\hat{\lambda}bq^{\prime}+\hat{F}\left(  q\right)
=-ab^{\frac{5}{2}}h^{\prime\prime}. \label{eq_two2}%
\end{equation}
For $\frac{\overline{q^{2}}}{\overline{g^{2}}}=\mathcal{B}_{0}$, we want to
find the set of parameters $\hat{\lambda}$ and $\hat{k}_{i}$ that will
maximize $P_{g}=\hat{\lambda}\overline{\dot{q}^{2}}$ given the dynamical
constraint (\ref{eq_one1}). This optimized quantity can also be written as%
\begin{equation}
P_{g}=\hat{\lambda}\overline{\dot{q}^{2}}=b^{2}\hat{\lambda}\overline
{q^{\prime2}}, \label{eq_two3}%
\end{equation}
where $q^{\prime}$ is described by the rescaled equation (\ref{eq_two2}).
However, the optimization problem in equations (\ref{eq_two2}) and
(\ref{eq_two3}) is identical with the original one given by equation
(\ref{eq_orgn}) and it has an optimal solution when $\hat{\lambda}%
=b\hat{\lambda}_{0}$ and $\hat{F}\left(  q\right)  =ab^{\frac{5}{2}}\hat
{F}_{0}\left(  \frac{q}{a\sqrt{b}}\right)  .$ For this set of parameters,
equation (\ref{eq_two2}) coincides with equation (\ref{eq_orgn}) and the
solution to (\ref{eq_two2}) will be $q\left(  t\right)  =a\sqrt{b}q_{0}\left(
bt\right)  $. Note that for this solution we also have%
\begin{equation}
\frac{\overline{q^{2}}}{\overline{g^{2}}}=\frac{a^{2}b\overline{q_{0}^{2}}%
}{a^{2}b\overline{h^{2}}}=\frac{\overline{q_{0}^{2}}}{\overline{h^{2}}%
}=\mathcal{B}_{0},
\end{equation}
and therefore the optimized solution that we found corresponds to the correct
response level. The last step is to compute the harvested power density for
the new solution. These will be given by%
\begin{equation}
\rho_{e,g}({\mathcal{B}_{0}})=\frac{\underset{\left\{  \hat{\lambda},\hat
{k}_{i}\text{ }|\text{ }\mathcal{B}_{0}\right\}  }{\max}\left(  \hat{\lambda
}\overline{\dot{q}^{2}}\right)  }{\omega_{g}^{3}\overline{q^{2}}}%
=\frac{\underset{\left\{  \hat{\lambda},\hat{k}_{i}\text{ }|\text{
}\mathcal{B}_{0}\right\}  }{\max}\left(  b^{2}\hat{\lambda}\overline
{q^{\prime2}}\right)  }{\left(  b^{3}\omega_{h}^{3}\right)  \left(
a^{2}b\overline{q_{0}^{2}}\right)  }=\frac{\left(  b\hat{\lambda}_{0}\right)
\left(  b^{3}a^{2}\overline{q_{0}^{\prime2}}\right)  }{\left(  b^{3}\omega
_{h}^{3}\right)  \left(  a^{2}b\overline{q_{0}^{2}}\right)  }=\frac
{\hat{\lambda}_{0}\overline{q_{0}^{\prime2}}}{\omega_{h}^{3}\overline
{q_{0}^{2}}}=\rho_{e,h}({\mathcal{B}_{0}}).
\end{equation}
This completes the proof.

We emphasize that the above property can be generalized for multi-dimensional
systems; a detailed study for this case will be presented elsewhere. {Through
this result we have illustrated that both uniform amplification and stretching
of the input spectrum }(see e.g. Figure \ref{fig_spectra} various amplified, and
stretched versions of the Pierson-Moskowitz) {will leave the harvested power
density unchanged, and therefore the shape of spectrum is the only factor  (i.e. relative
distribution of energy between harmonics) that modifies the harvested power density.}


Another important property of the developed measure is its independence of the
specific values of the system parameters since it always {refers} to the optimal
configuration for each design. Thus, it is an approach that characterizes a whole
class of systems rather than specific members of this class. To this end it is
suitable for the comparison of systems having different {forms} e.g. having
different function $\hat{F}\left(  q;\hat{k}_{1},...,\hat{k}_{n}\right)  $
since it is only the form of the system that is taken into account and not the
specific parameters $\hat{\lambda}$ and $\hat{k}_{1},...,\hat{k}_{n}$.

These two properties give an objective character to the derived measure as it
\textit{depends only on the form of the employed configuration and the form of
the input spectrum}. For this reason, it can be used to perform systematic
comparisons and optimizations among different classes of system
configurations, e.g. linear versus nonlinear harvesters. In addition to the
above properties, the curve $\rho_{e}\left(  \mathcal{B}\right)  $ reveals the
optimal response level $\overline{q^{2}}$ so that the harvested power over the
response magnitude is maximum, achieving {in this way} optimal utilization of
the device size.

We note that for a multi-dimensional energy harvester it may also be useful to
quantify the harvester performance using the\textit{ effective harvesting
coefficient} $\lambda_{e}$ which is defined as the maximum possible harvested
power $\underset{\left\{  \hat{\lambda},\hat{k}_{i}\text{ }|\text{
}\mathcal{B}\right\}  }{\max}P_{h}$ (for a given excitation spectrum and under
the {constraint} of a given response level $\mathcal{B}$) normalized by the total
kinetic energy of the device $E_{K}:$%
\begin{equation}
\lambda_{e}(\mathcal{B})=\frac{\underset{\left\{  \hat{\lambda},\hat{k}%
_{i}\text{ }|\text{ }\mathcal{B}\right\}  }{\max}P_{h}}{\omega_{h}E_{K}},
\end{equation}
where we have also non-dimensionalized with the mean frequency of the input
spectrum so that the ratio satisfies similar invariant properties under linear
transformations of the input spectrum. Although for MDOF systems the above
measure can provide useful information about the efficient utilization of
kinetic energy, for SDOF systems of the form (\ref{system1}) we always have
$\lambda_{e}(\mathcal{B})=\hat{\lambda}$ and to this end we will not study
this measure further in this work.

\section{Quantification of performance for SDOF harvesters}

We now apply the derived criteria in order to compare three different classes
of nonlinear SDOF energy harvesters excited by three qualitatively different
source spectra. In particular, we compare the performance of linear SDOF
harvesters with two classes of nonlinear oscillators: an essentially nonlinear
with cubic nonlinearity (mono-stable system) and one that has also cubic
nonlinearity but negative linear stiffness (double well potential system or
bistable) {as illustrated in Figure} \ref{fig_potential}. The first family of systems has been studied in various contexts
with main focus the improvement of the energy harvesting performance from
wide-band sources. The second family of nonlinear oscillators is well known
for its property to maintain constant vibration amplitudes even for very small
excitation levels, and it has also been applied to enhance the energy
harvesting capabilities of nonlinear energy harvesters. More specifically we
consider the following three classes of systems (Figure \ref{fig_sys}):%

\begin{figure}[H]
\centerline{
\begin{minipage}{\hsize}\begin{center}
\includegraphics[width=\hsize]{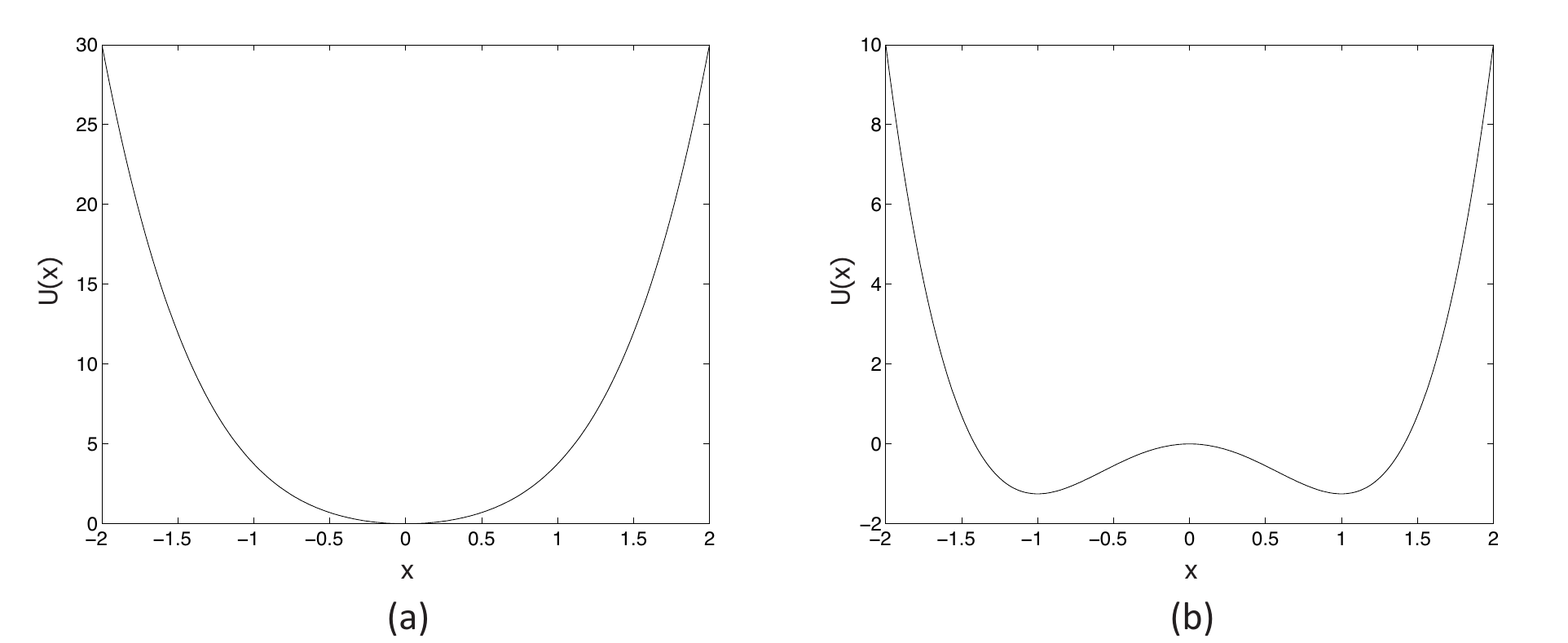}
\end{center}\end{minipage}
}
\caption{The shapes of potential function $U(x)=\frac{1}{2} \hat{k}_1 x^2 + \frac{1}{4} \hat{k}_3 x^4 $. (a) The monostable potential function with $\hat{k}_1>0$ and $\hat{k}_3>0$. (b) The bistable potential function with $\hat{k}_1<0$ and $\hat{k}_3>0$.    }
\label{fig_potential}
\end{figure}

\begin{align}
\ddot{q}+\hat{\lambda}\dot{q}+\hat{k}_1q  &  =-\ddot{h},\text{
\ \ \ \ \ \ \ \ \ \ \ \ \ \ \ \ \ \ (linear system)}\label{linear}\\
\ddot{q}+\hat{\lambda}\dot{q}+\hat{k}_3q^{3}  &  =-\ddot{h},\text{
\ \ \ \ \ \ \ \ \ \ \ \ \ \ \ \ \ \ (cubic system)}\label{cubic}\\
\ddot{q}+\hat{\lambda}\dot{q}-\hat{\nu}q+\hat{k}_3q^{3}  &  =-\ddot{h}.\text{
\ \ \ \ \ \ \ \ \ \ \ \ \ \ \ \ \ \ (negative stiffness)} \label{nega_stf}%
\end{align}
Our comparisons are presented for three cases of excitation spectra, namely:
the monochromatic excitation, the white noise excitation, and an intermediate
one characterized by colored noise excitation with Gaussian, stationary
probabilistic structure and a power spectrum having the Pierson-Moskowitz form%
\begin{equation}
S_{hh}=\frac{1}{\omega^{5}}\exp\left( - \omega^{-4}\right)  .
\label{PM_spectrum}%
\end{equation}%
\begin{figure}[ptb]%
\centering
\includegraphics[
height=1.8732in,
width=5.1655in
]%
{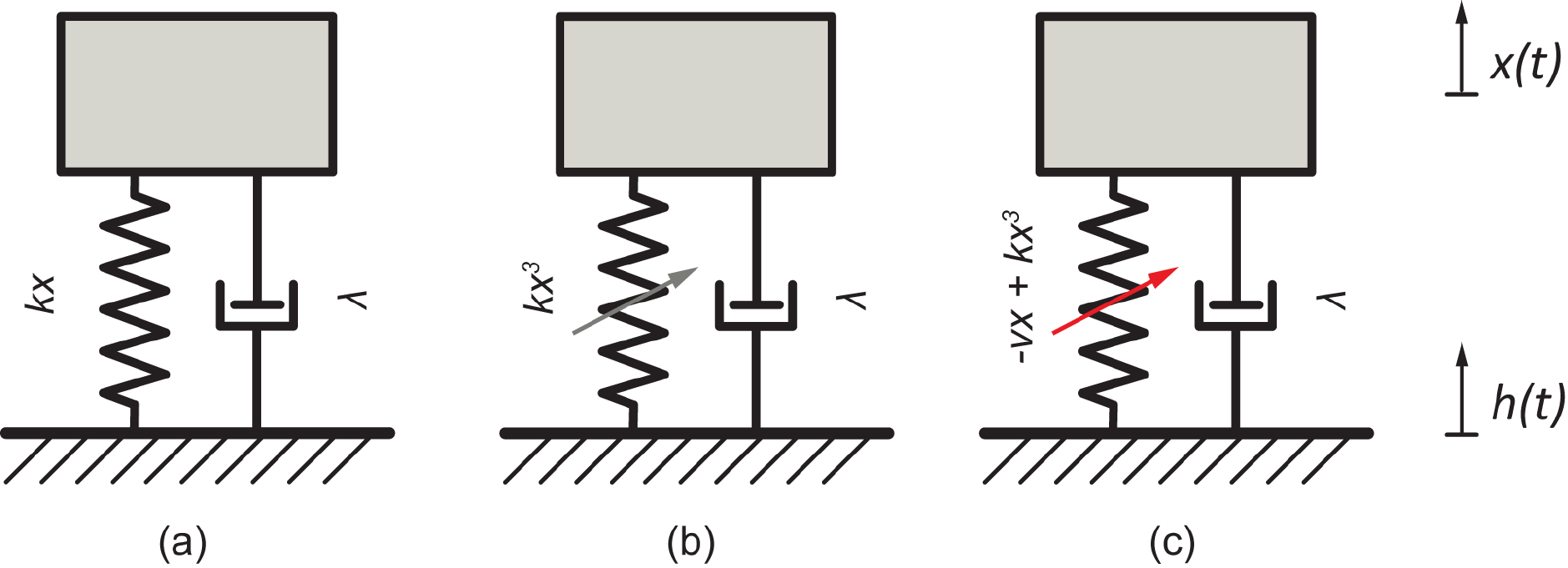}%
\caption{Linear and nonlinear SDOF systems: (a) Linear SDOF system, (b)
Nonlinear SDOF system only with a cubic spring, and (c) Nonlinear SDOF system
with the combination of a negative linear and a cubic spring.}%
\label{fig_sys}%
\end{figure}
The monochromatic and the white noise excitations are characterized by
diametrically opposed spectral properties: the first case is the extreme form
of a narrow-band excitation, while the second represents the most extreme case
of a wide-band excitation. Our goal is to understand and objectively compare
various designs that have been employed in the past to achieve better
performance from sources which are either monochromatic or broad-band. We are
also interested to use these two prototype forms of excitation in order to
interpret the behavior of SDOF harvesters for intermediate cases of excitation
such as the PM spectrum.

We first present the monochromatic and the white noise cases where many of the
results can be derived analytically. We analyze the critical differences in
terms of the harvester performance and subsequently, we numerically perform
stochastic optimization of the nonlinear designs for the intermediate PM
spectrum. For the PM excitation, we employ a discrete approximation of the
excitation $h$ in spectral space, with harmonics that have given amplitude but
relative phase differences modeled as uniformly distributed random variables.
The responses of the dynamical systems (\ref{linear}) and (\ref{cubic}) are
then characterized by averaging (after sufficient time so that transient
effects do not contribute) over a large ensemble of realizations, i.e.
averaging over a large number of excitations $h$ generated with a given
spectrum but randomly generated phases.

\subsection{SDOF harvester under monochromatic excitation}

\textbf{Linear system. }We calculate the harvested power density $\rho_{e}$
for the the linear oscillator under monochromatic excitation, i.e. the
one-sided power spectrum is given by $S_{hh}\left(  \omega\right)  =\alpha
^{2}\delta\left(  \omega-\omega_{0}\right)  $. For this case the computation
can be carried out analytically. In particular for the linear oscillator we
will have the power spectrum for the response given by%
\begin{equation}
S_{qq}\left(  \omega\right)  =\frac{\omega^{4}}{\left(  \hat{k}_1-\omega
^{2}\right)  ^{2}+\hat{\lambda}^{2}\omega^{2}}S_{hh}\left(  \omega\right) .
\label{eqn_spec_relation}%
\end{equation}
Thus, the response level can be computed as%
\begin{equation}
\mathcal{B}=\frac{\overline{q^{2}}}{\overline{h^{2}}}=\frac{\omega_{0}^{4}%
}{\left(  \hat{k}_1-\omega_{0}^{2}\right)  ^{2}+\hat{\lambda}^{2}\omega_{0}^{2}%
}, \label{eqn_LinMonoSpec}%
\end{equation}
where $\overline{h^{2}}$ is simply $\alpha^{2}$. Moreover, the average rate of
energy harvested per unit mass will be given by%
\begin{equation}
P_{h}=\hat{\lambda}\overline{\dot{q}^{2}}=\hat{\lambda}\alpha^{2}\frac
{\omega_{0}^{6}}{\left(  \hat{k}_1-\omega_{0}^{2}\right)  ^{2}+\hat{\lambda}%
^{2}\omega_{0}^{2}}.
\end{equation}
Then we will have from equation (\ref{eqn_LinMonoSpec})
\begin{equation}
\left(  \hat{k}_1-\omega_{0}^{2}\right)  ^{2}+\hat{\lambda}^{2}\omega_{0}%
^{2}=\frac{\omega_{0}^{4}}{\mathcal{B}}. \label{fig_inter_eq}%
\end{equation}
Thus, for a given $\mathcal{B}$, the mean rate of energy harvested will be
given by
\begin{equation}
P_{h}=\hat{\lambda}\overline{\dot{q}^{2}}=\hat{\lambda}\overline{q^{2}}%
\omega_{0}^{2}.
\end{equation}
Therefore the mean rate of energy harvested will become maximum when
$\hat{\lambda}$ is maximum. For fixed $\mathcal{B}$, equation
(\ref{fig_inter_eq}) shows that the maximum legitimate value of $\hat{\lambda
}$ will be given by $\hat{\lambda}=\frac{\omega_{0}}{\sqrt{\mathcal{B}}}$ and
this can be achieved when $\hat{k}_1=\omega_{0}^{2}.$ Therefore we will have
\begin{align}
P_{h}  &  =\frac{\omega_{0}^{3}\overline{q^{2}}}{\sqrt{\mathcal{B}}}%
=\omega_{0}^{3}\overline{h^{2}}\sqrt{\mathcal{B}}=\alpha^{2}\omega_{0}%
^{3}\sqrt{\mathcal{B}},\\
\rho_{e}  &  =\frac{\underset{\left\{  \hat{\lambda},\hat{k}_{i}\text{
}|\text{{}}\mathcal{B}\right\}  }{\max}P_{h}}{\omega_{h}^{3}\overline{q^{2}}%
}=\frac{1}{\sqrt{\mathcal{B}}}. \label{Linear_shp_monochr}%
\end{align}
Hence, for a linear SDOF system under monochromatic excitation, the harvested
power density is proportional to the magnitude of the square root of
$\mathcal{B}$ while the harvested power is proportional to the square root of
the response level.

\begin{figure}[ptb]%
\centering
\includegraphics[
height=6.7697in,
width=6.0295in
]%
{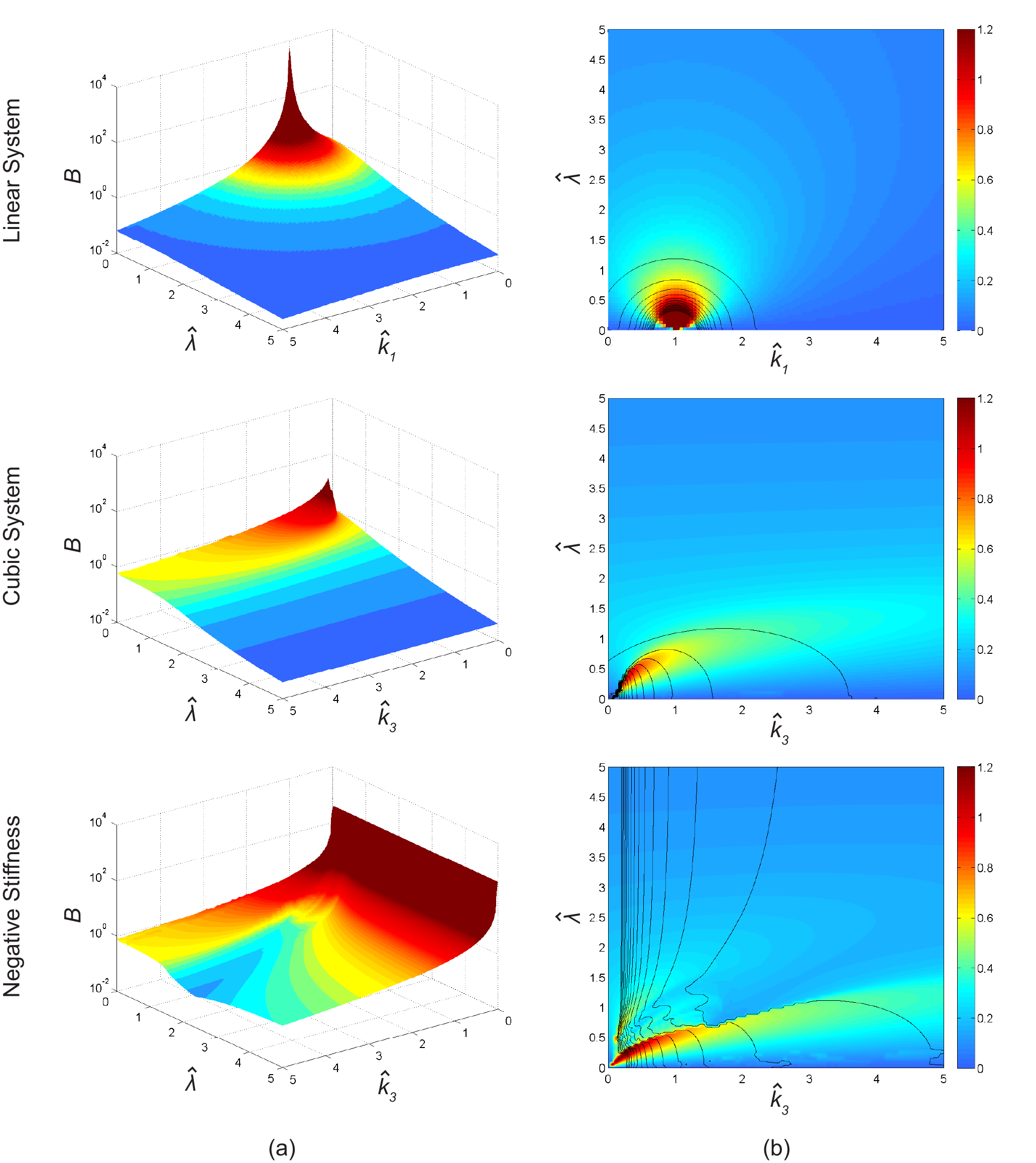}%
\caption{Response level $\mathcal{B}$ and power harvested for the case of
monochromatic spectrum excitation over different system parameters. The
response level $\mathcal{B}$ is also presented as a contour plot in the power
harvested plots. All three cases of systems are shown: linear (top row), cubic
(second row), and negative stiffness with $\hat{\nu}=1.$}%
\label{fig_mono_nonlinear}%
\end{figure}

\textbf{Cubic and negative stiffness harvesters. }For a nonlinear system the
response under monochromatic excitation cannot be obtained analytically and to
this end the computation will be carried out numerically. In figure
\ref{fig_mono_nonlinear},\ we present the response level $\mathcal{B}$ for all
three systems (linear, cubic, and the one with negative stiffness with
$\hat{\nu}=1$) for various system parameters. We also present the total
harvested power superimposed with contours of the response level $\mathcal{B}$.

\begin{figure}[ptb]%
\centering
\includegraphics[
height=2.3082in,
width=5.9819in
]%
{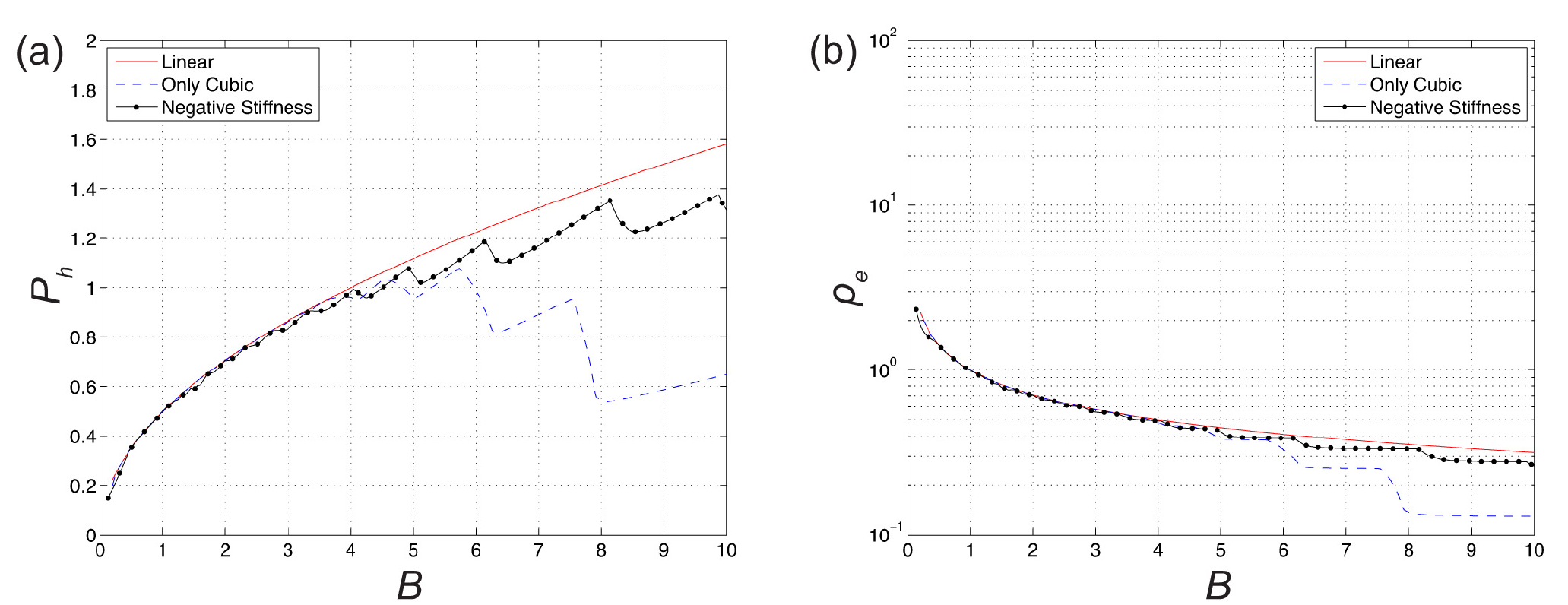}%
\caption{(a) Maximum harvested power, and (b) Power density for linear and
nonlinear SDOF systems under monochromatic excitation. }%
\label{fig_mono_shp}%
\end{figure}

For both the linear and the cubic oscillator, we can observe the 1:1 resonance
regime (see plots for the response level $\mathcal{B}$). For these two cases,
we also observe a similar decay of the response level with respect to the
damping coefficient. This behavior changes drastically in the negative
stiffness oscillator where the response level is maintained with respect to
changes of the damping coefficient. This is expected if one considers the
double well form of the corresponding potential that controls the amplitude of
the nonlinear oscillation. Despite the robust amplitude of the response, the
performance (i.e. the amount of power being harvested) drops similarly with
the other two oscillators (especially the cubic one) as the damping
coefficient increases. Therefore robust response level does not necessarily imply constant performance level. %
To quantify the performance,\ we present in Figure \ref{fig_mono_shp} the
maximum harvested power and the harvested power density for the three
different oscillators. We observe that in all cases the linear design has
superior performance compared with the nonlinear configurations. {In addition,
we note that the cubic and the negative stiffness oscillators have strongly
variable performance which presents non-monotonic behavior with respect to the
response level $\mathcal{B}$.} 

To better understand the nature of this variability, we pick two characteristic
values of $\mathcal{B}$ (one close to a local minimum i.e. $\mathcal{B=}8.5$
and one at a local maximum, i.e. $\mathcal{B=}8.1$) for the negative stiffness
oscillator (Figure \ref{fig_mono_shp})$.$ From
these points, we can observe that the strong performance for the nonlinear
oscillator is associated with signatures of 1:3 resonance in the response
spectrum. We also note that the small amplitude of the higher harmonic is not
sufficiently large to justify the difference in the performance. On the other
hand, the significant amplitude difference on the primary harmonic, which can
be considered as an indirect effect of the 1:3 resonance, justifies the strong
variability between the two cases.%
\begin{figure}[ptb]%
\centering
\includegraphics[
height=2.3073in,
width=5.9127in
]%
{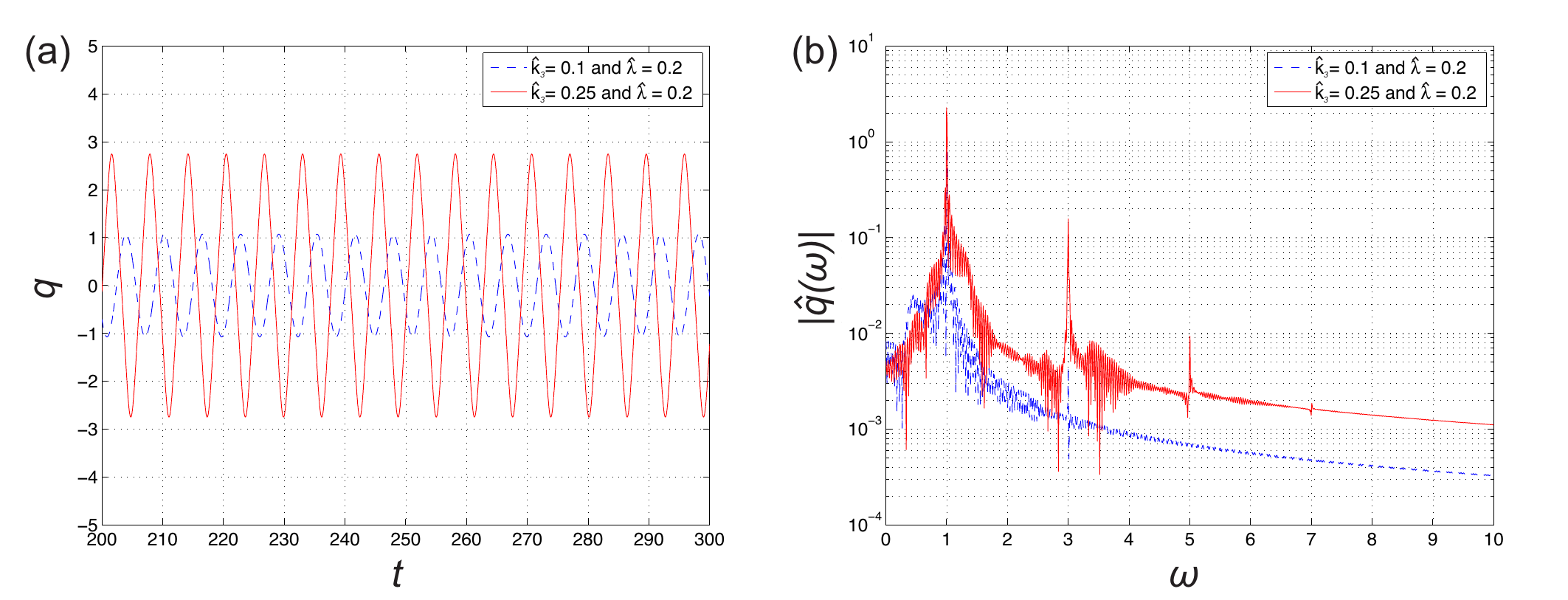}%
\caption{A nonlinear system with the combination of a negative linear
($\hat{\nu}=1$) and a cubic spring. Blue solid line corresponds to a local
minimum of the performance in Fig. \ref{fig_mono_nonlinear}: $\hat{k}_3=0.1$ and
$\hat{\lambda}=0.2$. Red dashed line corresponds to a local maximum of the
performance in Fig. \ref{fig_mono_nonlinear}: $\hat{k}_3=0.25$ and $\hat{\lambda}=0.2$. (a)
Response in terms of displacement. (b) Fourier transform modulus $\left\vert
\hat{q}\left(  \omega\right)  \right\vert $.}%
\label{fig_nega_resp}%
\end{figure}

Independently of the super-harmonic resonance occurring in the nonlinear
designs for certain response levels, it is clear that the best performance for
SDOF systems under monochromatic excitation can be achieved within the class
of linear harvesters. To understand this result, we consider the general
equation (\ref{Eq_org_trans}) multiplying with $\dot{q}$ and applying the mean
value operator. This will give us the following energy equation%
\begin{equation}
\frac{1}{2}\frac{d}{dt}\left(  \overline{\dot{q}^{2}}\right)  +\hat{\lambda
}\overline{\dot{q}^{2}}+\overline{\hat{F}\left(  q\right)  \dot{q}}%
=-\overline{\ddot{h}\dot{q}}.
\end{equation}
In a statistical steady state, we will have the first term vanishing. This is
also the case for the third term, which represents the overall energy
contribution from the conservative spring force. Moreover, the harvested power
is equal to the second term and thus we have%
\begin{equation}
P_{h}=\hat{\lambda}\overline{\dot{q}^{2}}=-\overline{\ddot{h}\dot{q}}.
\end{equation}
For the monochromatic case, we have $\ddot{h}\left(  t\right)  =-\alpha
\omega_{0}^{2}\cos\omega_{0}t.$ We represent the arbitrary statistical steady
state response as%
\begin{equation}
q=%
{\displaystyle\sum\limits_{i}}
\hat{q}_{i}\cos\left(  \omega_{i}t+\phi_{i}\right) ,
\end{equation}
with $\hat{q}_{i}>0$, and { $\phi_{i}$ are phases determined from the system dynamics}. From this representation, we obtain%
\begin{equation}
P_{h}=%
{\displaystyle\sum\limits_{i}}
\hat{q}_{i}\alpha\omega_{0}^{2}\omega_{i}\lim_{T\rightarrow\infty}\frac{1}{T}%
{\displaystyle\int\limits_{0}^{T}}
\cos\omega_{0}t\sin\left(  \omega_{i}t+\phi_{i}\right)  dt.
\end{equation}
The quantity inside the integral will be nonzero only when $i=0$. Thus,%

\begin{equation}
P_{h}=\hat{q}_{0}\alpha\omega_{0}^{3}\frac{\omega_{0}}{2\pi}%
{\displaystyle\int\limits_{0}^{\frac{2\pi}{\omega_{0}}}}
\cos\omega_{0}t\sin\left(  \omega_{0}t+\phi_{0}\right)  dt=\frac{1}{2}%
\hat{q}_{0}\alpha\omega_{0}^{3}\sin\phi_{0}.
\end{equation}

Note that from the representation for $q$, we obtain
\begin{align}
\overline{q^{2}}  &  =%
{\displaystyle\sum\limits_{i,j}}
\hat{q}_{i}\hat{q}_{j}\overline{\cos\left(  \omega_{i}t+\phi_{i}\right)
\cos\left(  \omega_{j}t+\phi_{j}\right)  }\nonumber\\
&= \sum_{i, j (i\neq j)} \frac{1}{2} \hat{q}_i\hat{q}_j \overline{\{\cos ([\omega_i-\omega_j]t + \phi_i-\phi_j) + \cos ([\omega_i+\omega_j]t + \phi_i+\phi_j)\}} \nonumber \\ 
&\ \ + \sum_{i} \frac{1}{2} \hat{q}_i^2 \overline{\{1+ \cos (2\omega_it + 2\phi_i)\}} \nonumber \\
&=\frac{1}{2}%
{\displaystyle\sum\limits_{i}}
\hat{q}_{i}^{2}.
\end{align}
It is straightforward to conclude that for constant response level
$\overline{q^{2}}$ the harvested power will become maximum when $\hat{q}_{0}$
is maximum, and this is the case only when all the energy of the response is
concentrated in the harmonic $\omega_{0}$, a property that is guaranteed to
occur for the linear systems.\ \textit{Thus, for SDOF harvesters, excited by
monochromatic sources, the optimal linear system can be considered as an upper
bound of the performance among the class of both linear and nonlinear
oscillators.}

\subsection{SDOF harvester under white noise excitation}

We investigated the monochromatic excitation case of both linear and nonlinear
systems as an extreme case of a narrow-band excitation. The opposite extreme,
the one that corresponds to a broadband excitation, is the Gaussian white
noise. We consider a dynamical system governed by a second order differential
equation under the standard Gaussian white noise excitation $\dot{W}(t)$ with
zero mean and intensity equal to one (i.e. $\overline{W^{2}}=1$).%
\begin{equation}
\ddot{q}+\hat{\lambda}\dot{q}+\hat{F}(q)=\alpha\dot{W}(t).
\label{SDOF1_white}%
\end{equation}
For this SDOF system, the probability density function is fully described by
the Fokker-Planck-Kolmogorov equation which for the statistical steady state
can be solved analytically providing us with the exact statistical response of
system (\ref{SDOF1_white}) in terms of the steady state probability density
function (see e.g. \cite{sobczyk91})%
\begin{equation}
p_{st}\left(  q,\dot{q}\right)  =C\exp\left(  {-\frac{\hat{\lambda}}%
{\alpha^{2}}\left[  \frac{\dot{q}^{2}}{2}+\int_{0}^{q}\hat{F}(x)dx\right]
}\right) , \label{pdf_white}%
\end{equation}
where $C$ is the normalization constant so that $%
{\displaystyle\iint}
p_{st}\left(  q,\dot{q}\right)  dqd\dot{q}=1.$

\begin{figure}[ptb]%
\centering
\includegraphics[
height=2.3834in,
width=6.0027in
]%
{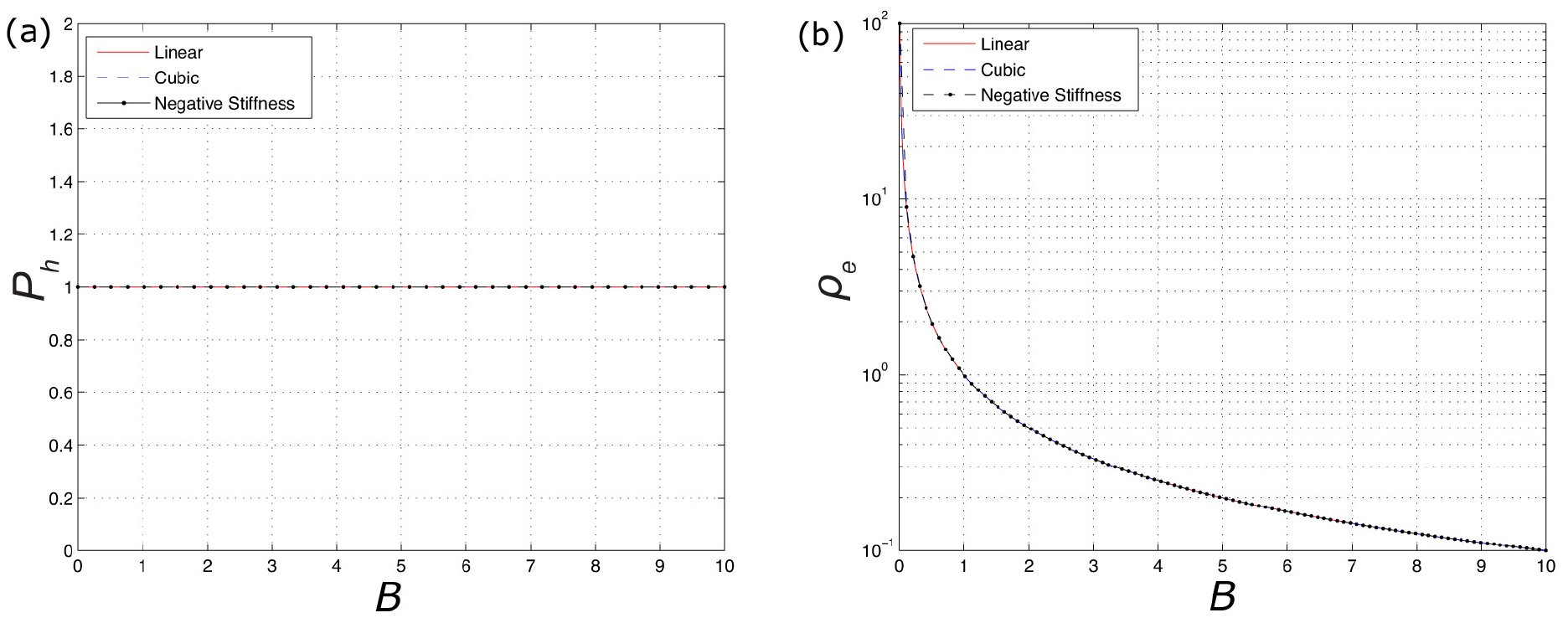}%
\caption{(a) Maximum harvested power, and (b) Power denstity for linear and
nonlinear SDOF systems under white noise excitation. }%
\label{fig_white}
\end{figure}

In order to use previously developed measures, we define $\overline{h^{2}%
}=\alpha^{2}$ (the typical amplitude of the excitation is equal to the
intensity of the noise). Moreover, since there is no characteristic frequency
we can choose without loss of generality $\omega_{h}^{2}=1.$ Using expression
(\ref{pdf_white}), we can compute an exact expression for the harvested power%
\begin{equation}
P_{\dot{W}}=\hat{\lambda}\overline{\dot{q}^{2}}=\alpha^{2}.
\end{equation}
which is an independent quantity of the system parameters - the above result
can be generalized in MDOF system as shown in \cite{LangleyRS}. We observe
that in this extreme form of broadband excitation the harvested power is
independent on the system parameters and depends only on the excitation energy
level $\alpha.$ In addition, the harvested power density $\rho_{e}$ will be
given by%
\begin{equation}
\rho_{e}(\mathcal{B})=\frac{\underset{\left\{  \hat{\lambda},\hat{k}_{i}\text{
}|\text{ }\mathcal{B}\right\}  }{\max}P_{h}}{\omega_{h}^{3}\overline{q^{2}}%
}=\frac{\alpha^{2}}{\overline{q^{2}}}=\frac{1}{\frac{\overline{q^{2}}%
}{\overline{h^{2}}}}=\frac{1}{\mathcal{B}}.
\end{equation}
Similarly with the harvested power, we observe that the harvested power
density is also independent {of} the employed system design (Figure
\ref{fig_white}). Moreover, when we compare with the monochromatic excitation
case (where we illustrated that the best possible performance can be achieved
with linear systems), we see that the harvested power density drops faster
with respect to the device size $\mathcal{B}$ when the energy is spread (in
the spectral sense) compared with the case where energy is localized in a
single input frequency.%

\subsection{SDOF harvester under colored noise excitation}

\begin{figure}[ptb]%
\centering
\includegraphics[
height=6.4601in,
width=5.5694in
]%
{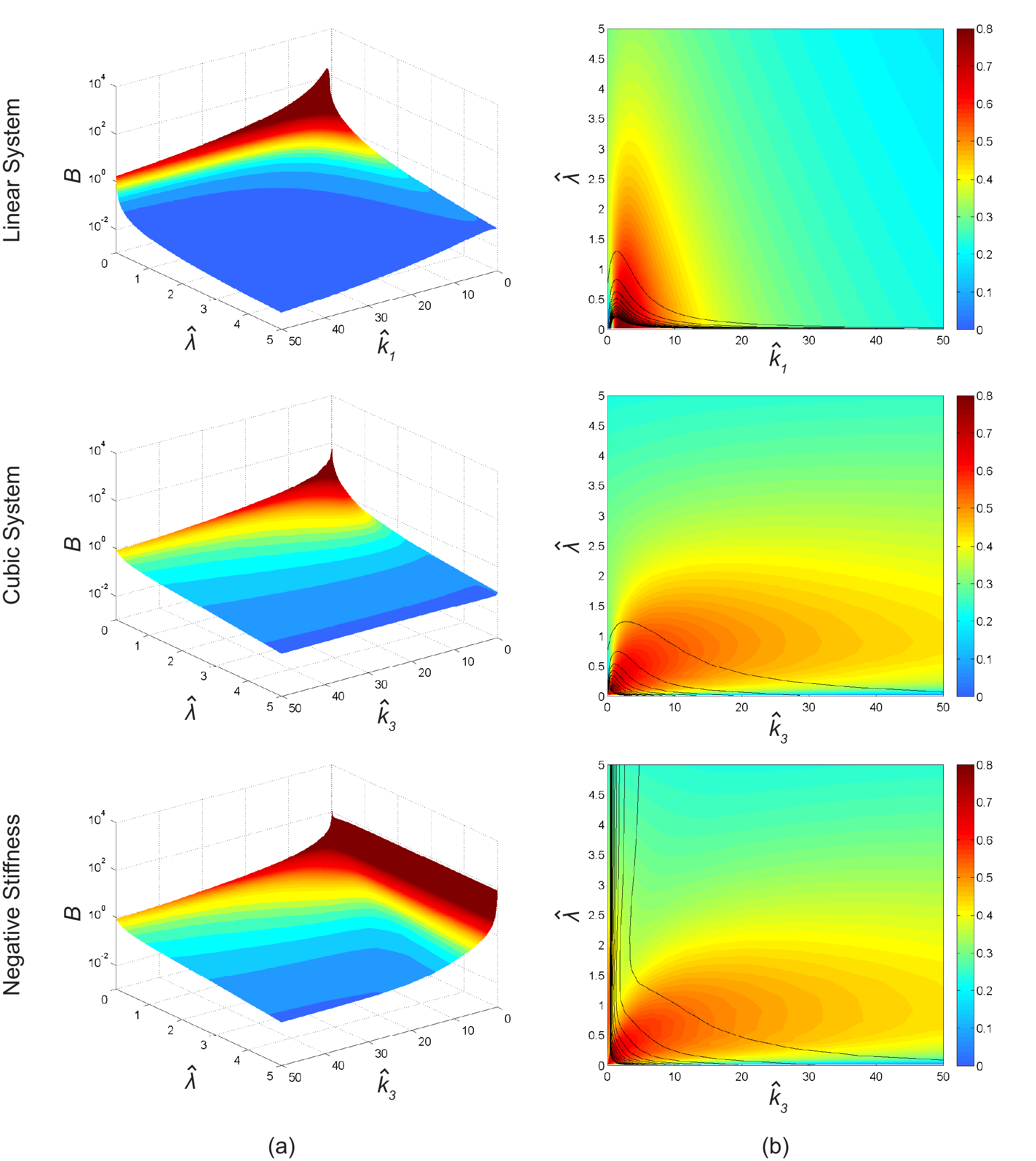}%
\caption{Response level $\mathcal{B}$ and power harvested for the case of
excitation with Pierson-Moskowitz spectrum over different system parameters.
The response level $\mathcal{B}$ is also presented as a contour plot in the
power harvested plots. All three cases of systems are shown: linear (top row),
cubic (second row), and negative stiffness with $\hat{\nu}=1.$}%
\label{nonlin_PM}%
\end{figure}

The third case of our analysis involves a colored noise excitation, the
Pierson-Moskowitz form (equation \ref{PM_spectrum}), which can be considered
as an intermediate case between the two extremes presented previously. For a
general excitation spectrum, the computation of the performance measures for
the nonlinear systems has to be carried out numerically. However for the
linear system the computation of the mean square amplitude and the mean rate
of energy harvested per unit mass can be computed analytically
\cite{sobczyk91}
\begin{align}
\overline{q^{2}}\left(  \hat{k}_1,\hat{\lambda}\right)   &  =%
{\displaystyle\int\limits_{0}^{\infty}}
\frac{\omega^{4}}{\left(  \hat{k}_1-\omega^{2}\right)  ^{2}+\hat{\lambda}%
^{2}\omega^{2}}\frac{1}{\omega^{5}}\exp\left(  -\omega^{-4}\right)  d\omega,\\
P_{h}\left(  \hat{k}_1,\hat{\lambda}\right)   &  =\hat{\lambda}%
{\displaystyle\int\limits_{0}^{\infty}}
\frac{\omega^{6}}{\left(  \hat{k}_1-\omega^{2}\right)  ^{2}+\hat{\lambda}%
^{2}\omega^{2}}\frac{1}{\omega^{5}}\exp\left(  -\omega^{-4}\right)  d\omega.
\end{align}
For the nonlinear systems, we employ a Monte-Carlo method since the
computational cost for simulating the SDOF harvester is reasonable. In
particular, we generate random realizations which are consistent with the PM
spectrum using a frequency domain method \cite{percival92}. Subsequently, we simulate the dynamics of the SDOF long enough so that the system reaches a statistical steady state. 
The results are
presented in Figure \ref{nonlin_PM}. We can still observe similar features
with the monochromatic excitation even though the variations of response level
and performance are now much smoother (compared with the monochromatic case).
For the linear system, we do not have the sharp resonance peak that we had in
the monochromatic case while the two nonlinear designs behave very similarly
in terms of their performance maps. However, the characteristic difference of
the negative stiffness design, related to the persistence of the response
level even for large values of damping, is preserved in this non-monochromatic
excitation case. Note that similarly to the monochromatic case this robustness
in the response level does not necessarily imply strong harvesting power.%

\subsection{Comparison of the three different systems}

\begin{figure}[ptb]%
\centering
\includegraphics[
height=2.092in,
width=5.8185in
]%
{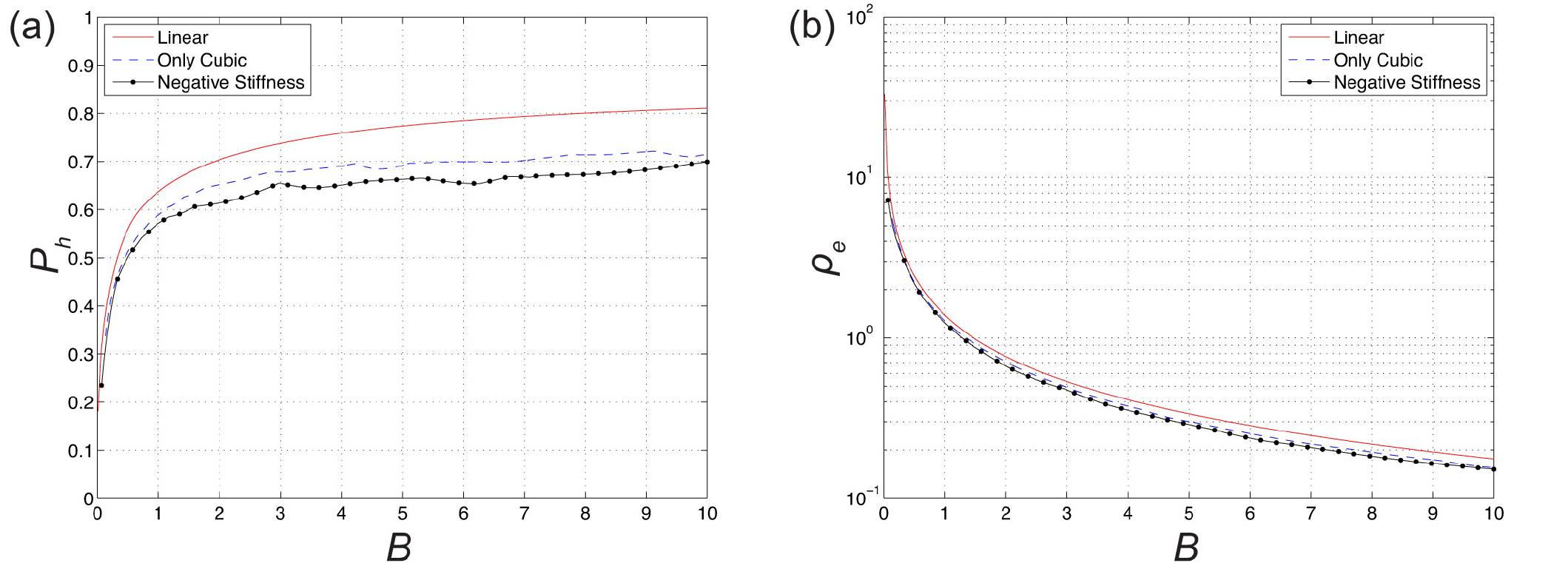}%
\caption{(a) Maximum harvested power, and (b) Power density for linear and
nonlinear SDOF systems under Pierson-Moskowitz spectrum.}%
\label{comp_PM}%
\end{figure}

A comparison of the linear system and the nonlinear systems under the
Pierson-Moskowitz spectrum excitation is shown in Figure \ref{comp_PM}. As it
can be seen from Figure \ref{comp_PM}b, the linear oscillator has the best
performance compared to two nonlinear designs (note that for the negative
stiffness oscillator a wide range of values $\hat{\nu}$ was employed and in
all cases the results for the power density were qualitatively the same - to
this end only the case $\hat{\nu}=1$ is presented). This is expected for any
colored noise excitation, given that for the monochromatic extreme we have
shown rigorously that the optimal performance of any nonlinear oscillator
cannot exceed the optimal linear design, while for the white noise excitation
all designs have identical performance. The relative performance in terms of harvested power for the three different classes of oscillators under three different types of excitations are summarized in Table \ref{table_comp}.
\begin{table}[H]
  \centering
  \caption{Harvested power of three different classes of oscillators under three different types of excitations. $\mathcal{L}$ denotes the linear oscillator, $\mathcal{M_N}$ denotes the monostable nonlinear oscillator, and $\mathcal{B_N}$ denotes the bistable nonlinear oscillator. } 
  \vspace{.1in}
  \begin{tabular}{M| c}
    \toprule
    Excitation & Performance comparison  \\
    \midrule     \midrule
Monochromatic excitation & $\mathcal{L} >>\mathcal{B_N} >\mathcal{M_N} $  \\ 
    \midrule
    Colored  noise excitation & $\mathcal{L} >\mathcal{M_N} >\mathcal{B_N} $\\ 
    \midrule
    White noise excitation & $\mathcal{L} = \mathcal{M_N}  = \mathcal{B_N} $ \\ 
    \bottomrule
  \end{tabular}
    \label{table_comp} 
\end{table}

\begin{figure}[ptb]%
\centering
\includegraphics[
height=2.6705in,
width=3.5596in
]%
{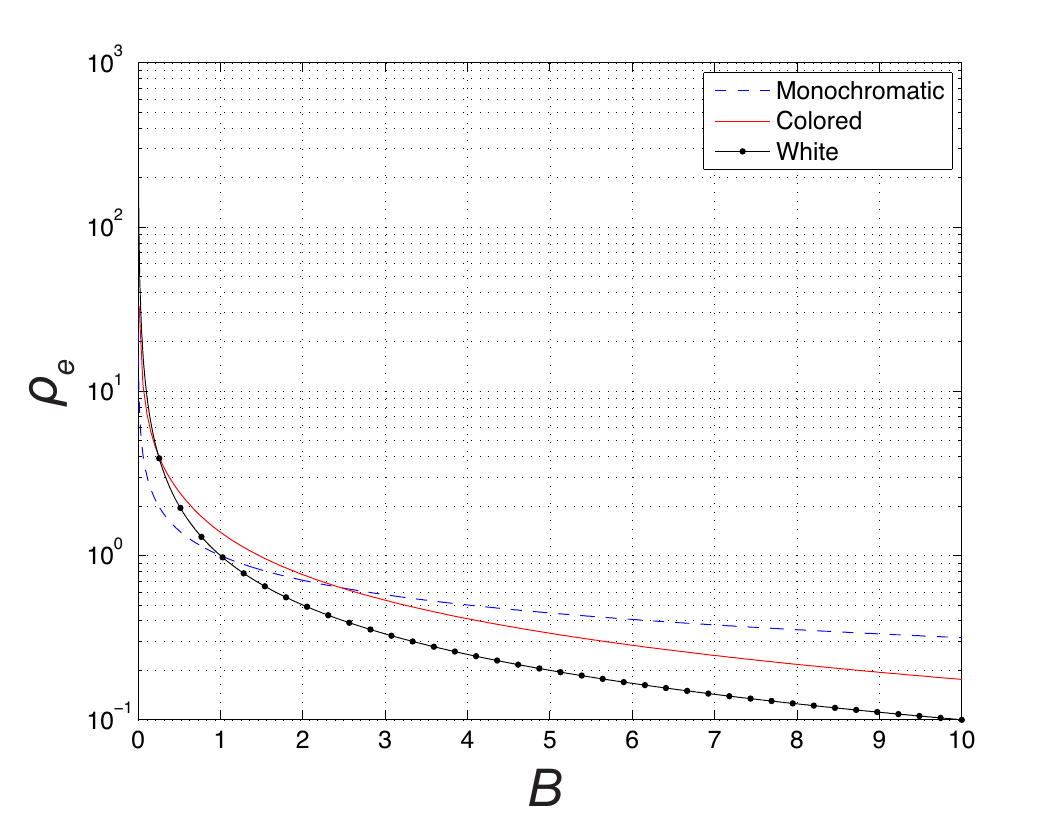}%
\caption{Harvested power density $\rho_{e}$ for the three different types of
excitation spectra. The linear design is used in all cases since this is the
optimal.}%
\label{shp_comapre_spectra}%
\end{figure}

An important qualitative difference between the response under the
Pierson-Moskowitz spectrum and the monochromatic excitation is the behavior of
the harvested power for larger values of $\mathcal{B}$. While for the
monochromatic case the harvested power scales with $\sqrt{\mathcal{B}}$, this
is not the case for the colored noise excitation where the harvested power
seems to converge to a finite value (a behavior that is consistent with the
white noise excitation). Therefore, we can conclude that for small values of
response level $\mathcal{B}$ the optimal performance under colored noise
excitation behaves similarly with the monochromatic excitation while for
larger values of $\mathcal{B}$ the optimal performance seems to be closer to
the white-noise response. The above conclusions are also verified from Figure
\ref{shp_comapre_spectra} where the three optimal harvested power density
curves (corresponding to the three forms of excitation) are presented
together. The performance of the optimal SDOF energy harvesters with respect to the size of the device $\mathcal{B}$ is incorporated in Table \ref{table_ph}.
\begin{table}[ht]
  \centering
  \caption{The performance of  optimal SDOF energy harvesters with respect to the size of the device. The number of $+$ indicates under which excitation the energy harvester performs relatively better.} 
  \vspace{.1in}
  \begin{tabular}{M| c|c|c}
    \toprule
    Case & Monochromatic & Colored noise & White noise \\
    \midrule     \midrule
$\mathcal{B}<1$ &$+$  &$++$ &$+++$\\ 
    \midrule
$\mathcal{B}>1$&$+++$  &$++$ &$+$\\ 
    \midrule
$\mathcal{B}=1$&$+$  &$+++$ &$+$\\ 
    \bottomrule
  \end{tabular}
  \label{table_ph} 
\end{table}

%

\section{Quantification of performance robustness}

We have examined the optimal performance for different designs of SDOF
harvesters under various forms of random excitations. Even though the linear
design has the optimal performance for fixed response level $\mathcal{B}$, the
robustness of this performance under perturbations of the input spectrum
characteristics (and with fixed optimal system parameters) has not been
considered. This is the scope of this section where we investigate how linear
and nonlinear systems with optimal system parameters behave when the
excitation spectrum is perturbed.

More specifically, we are interested to investigate robustness properties with
respect to frequency shifts of the excitation spectrum. Clearly, the harvested
power and the response level (that characterizes the size of the device) will
be affected by the spectrum shift. To quantify these variations we consider
the following three ratios%
\begin{equation}
\delta=\frac{B_{shifted}}{B_{0}},\text{ \ \ \ \ }\tau=\frac{(P_{h})_{shifted}%
}{(P_{h})_{0}},\text{\ \ \ \ \ \ }\sigma=\frac{(\rho_{e})_{shifted}}{(\rho
_{e})_{0}},
\end{equation}
where $\delta$ quantifies the variation of the response level ${\mathcal{B}%
}_{0}$ which essentially expresses the size of the device, $\tau$ quantifies
exclusively the changes in performance while $\sigma$ shows the changes in
harvested power density, i.e. it also takes into account the variations of the
response level ${\mathcal{B}}$.

\begin{figure}[ptb]%
\centering
\includegraphics[
height=4.702in,
width=6.007in
]%
{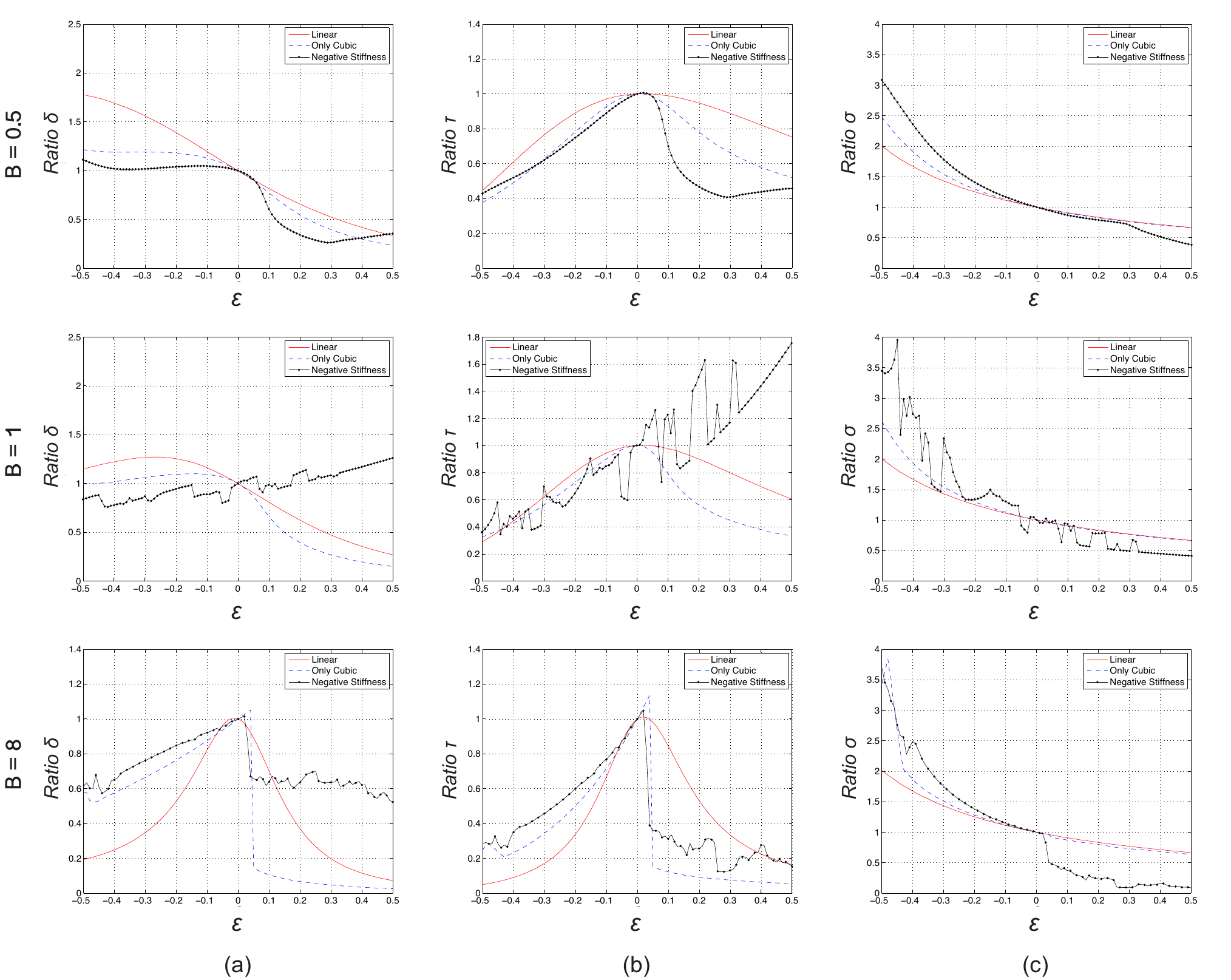}%
\caption{Robustness of (a) the response level, (b) the power harvested,
and (c) the harvested power density for the
\textit{monochromatic excitation} under three regimes of operation:
$\mathcal{B}=0.5$, $\mathcal{B}=1$, and $\mathcal{B}=8$.}%
\label{robust_mono}%
\end{figure}

\textbf{Monochromatic excitation.} For the monochromatic excitation,
perturbation in terms of spectrum shift can be expressed as
\begin{equation}
S_{hh}(\omega-\epsilon)=\delta(\omega-\omega_{0}-\epsilon).
\end{equation}
where $\omega_{0}=1$. In Figure \ref{robust_mono}, we present the ratios
describing the variation of the response level $\delta$, the harvested power
$\tau$ and the harvested power density $\sigma$ in terms of perturbation
$\epsilon$ for various levels of the unperturbed response level ${\mathcal{B}%
}$. For small response {levels}, i.e. when the system response is smaller than
the excitation (${\mathcal{B=}}0.5$) we observe that the negative stiffness
oscillator has more robustness {to} maintaining its response level when it is
excited by {lower} frequencies $\left(  \epsilon<0\right)  $. For the same
case, the harvested power decays in a similar fashion with the other two
oscillators. Therefore, for $\epsilon<0$ and ${\mathcal{B=}}0.5$ the nonlinear
oscillator with negative stiffness has the most robust performance. For faster
excitations $\left(  \epsilon>0\right)  $ we observe that all oscillators drop
their response level in smaller values than the design response level
${\mathcal{B}}_{0}$ with the linear system having the most robust behavior in
terms of the total harvested power. We emphasize that as long as $\delta<1$
robustness is essentially defined by the largest value of $\tau$ among
different types of oscillators.

For ${\mathcal{B=}}1$, we can observe that for all values of $\epsilon$ the
negative stiffness oscillator has the most robust behavior in terms of the
excitation level while the behavior of the harvested power is also better
compared with the other two classes of oscillators. For larger values of the
response level (${\mathcal{B=}}8$), we note that the response level ratio
$\delta$ is maintained in levels below 1; therefore the size of the device
will not be exceeded due to input spectrum shifts. On the other hand when we
consider the variations of the harvested power, we observe that all in all the
linear oscillators has the most robust behavior, while the two nonlinear
oscillators drop suddenly their performance to very small levels for larger,
positive values of $\epsilon.$%

\begin{figure}[ptb]%
\centering
\includegraphics[
height=4.6198in,
width=6.1151in
]%
{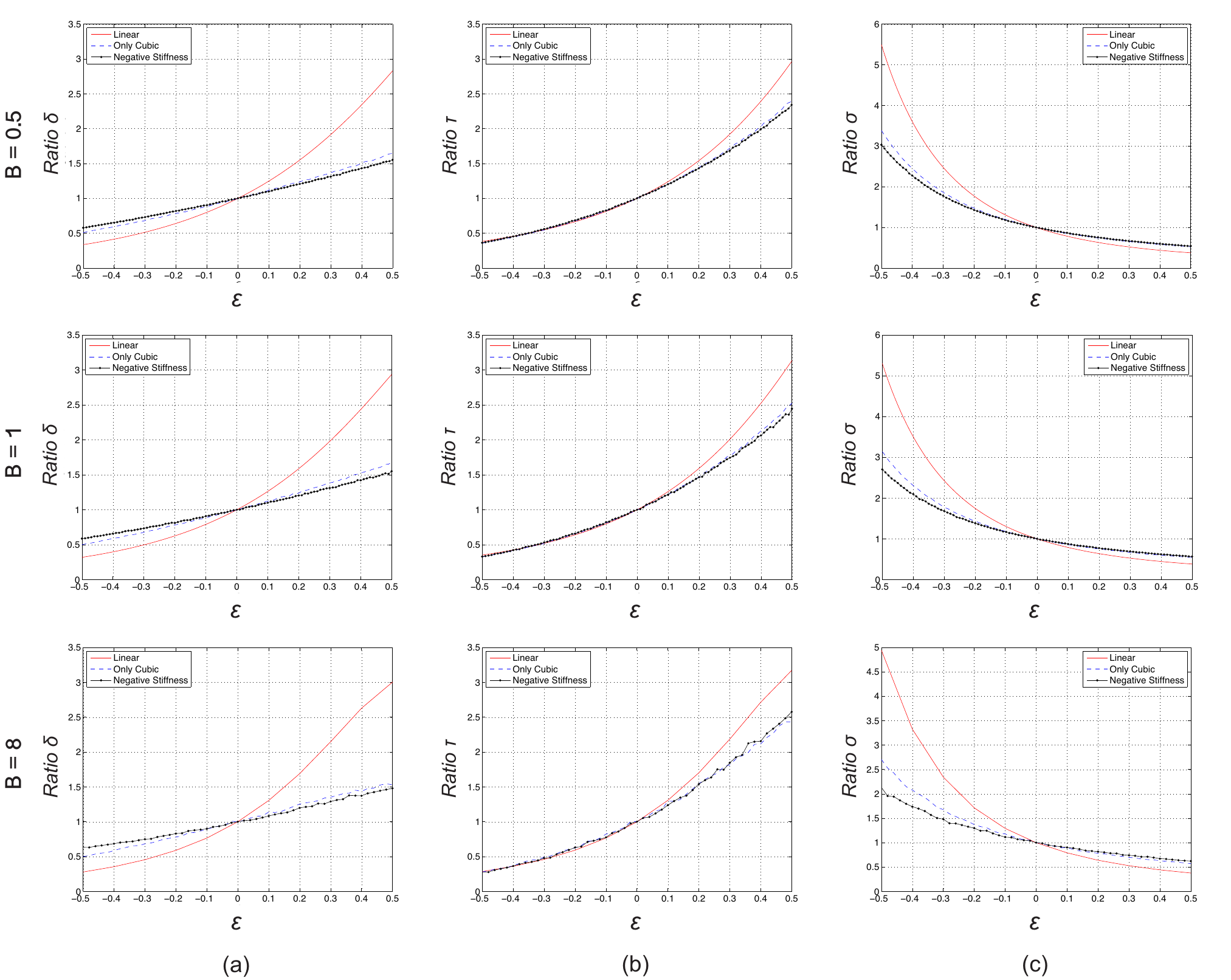}%
\caption{Robustness of (a) the response level, (b) the power harvested,
and (c) the harvested power density for the
\textit{PM spectrum excitation} under three regimes of operation:
$\mathcal{B}=0.5$, $\mathcal{B}=1$, and $\mathcal{B}=8$.}%
\label{robust_color}%
\end{figure}

\textbf{Colored noise excitation.} Similarly with the monochromatic case, we
consider a small perturbation $\epsilon$ for the colored noise excitation
spectrum:%
\begin{equation}
S_{hh}(\omega-\epsilon)=\frac{1}{(\omega-\epsilon)^{5}}\exp\left(
-(\omega-\epsilon)^{-4}\right)  .
\end{equation}
The results are presented in Figure \ref{robust_color} for three different
cases of unperturbed excitation levels ${\mathcal{B}}_{0}$. In contrast to the
monochromatic case, the ratios $\delta,\tau,$ and $\sigma$ have much {smoother}
dependence on the perturbation $\epsilon.$ Moreover, their variation is very
similar for all three response levels ${\mathcal{B}}_{0}$. More specifically,
we can clearly see that the two classes of nonlinear oscillators can better
maintain their response level over all values of $\epsilon.$ On the other
hand, the linear oscillator obtains a larger response level ${\mathcal{B}}$
when the spectrum is shifted to the right $\left(  \epsilon>0\right)  $
without substantially increasing the harvested power compared with the other
two nonlinear oscillators. For $\epsilon<0$, all three families of oscillators
harvest the same amount of energy. Thus, for colored noise excitation, the two
families of nonlinear oscillators achieve the most robust performance. Hence,
as long as the nonlinear design is chosen so that it has comparable optimal
performance with the family of linear oscillators, it is the preferable choice
since it has the best robustness properties.

\section{Conclusions}

We have considered the problem of energy harvesting using SDOF oscillators. We
first developed objective measures that quantify the performance of general
nonlinear systems from broadband spectra, i.e. simultaneous excitation from a
broad range of harmonics. These measures explicitly take into account the
required size of the device in order to achieve this performance. We
demonstrated that these measures do not depend on the magnitude or the
temporal scale of the input spectrum but only the relative distribution of
energy among different harmonics. In addition they are suitable to compare
whole classes of oscillators since they always pick the most effective
{parameter} configuration.

Using analytical and numerical methods, we applied the developed measures to
quantify the performance of three different families of oscillators (linear,
essentially cubic, and negative stiffness or bistable) for three different
types of excitation spectra: an extreme form of a narrow band excitation
(monochromatic excitation), an extreme form of a wide-band excitation
(white-noise), and an intermediate case involving colored noise
(Pierson-Moskowitz spectrum). For all three cases, we presented numerical and
analytical arguments that the nonlinear oscillators can achieve in the best
case equal performance with the optimal linear oscillator, given that the size
of the device does not change. We also considered the robustness of each
design to input spectrum shifts concluding that the nonlinear oscillator has
the best behavior for the colored noise excitation. {To this end, we concluded that, under a situation of designing a harvester with specific power, a nonlinear oscillator 
designed to achieve a performance that is close to
the optimal performance of a linear oscillator is the best choice since it
also has robustness against small perturbations.}


Future work involves the generalization of the presented criteria to MDOF
oscillators and the study of the benefits due to nonlinear energy transfers
between modes \cite{sapsis_et_al09, sapsis_PEM11, sapsis_JVA12,
Vakakis_book_08}. Preliminary results indicate that the application of
nonlinear energy transfer ideas can have a significant impact on achieving
higher harvested power density by distributing energy to more than one modes
achieving in this way smaller required device size without reducing its
performance level.

\textbf{Acknowledgments.} The authors would like to acknowledge the support
from Kwanjeong Educational Foundation as well as a startup grant at MIT. TPS
is also grateful to the American Bureau of Shipping for support under a Career
Development Chair.

\section*{Appendix I: An overview of spectral properties for stationary and
ergodic signals}

Here we recall some basic properties for random signals. Let $h\left(
t\right)  $ be a \textit{stationary} and \textit{ergodic} signal for which we
assume that it has finite power, i.e.
\[
\lim_{T\rightarrow\infty}\frac{1}{2T}%
{\displaystyle\int\limits_{-T}^{T}}
\left\vert h\left(  t\right)  \right\vert ^{2}dt<\infty.
\]
We define the \textit{correlation function}%
\[
R_{hh}\left(  \tau\right)  =\lim_{T\rightarrow\infty}\frac{1}{2T}%
{\displaystyle\int\limits_{-T}^{T}}
h\left(  t\right)  h\left(  t+\tau\right)  dt=\overline{h\left(  t\right)
h\left(  t+\tau\right)  },
\]
where the bar denotes ensemble averaging and the last equality follows from
the assumption of ergodicity. Note that we always have the property
\[
\left\vert R_{hh}\left(  \tau\right)  \right\vert \leq R_{hh}\left(  0\right)
.
\]
Based on the correlation function, we can compute the \textit{power spectrum}%
\[
S_{hh}\left(  \omega\right)  =\mathcal{F}\left[  R_{hh}\left(  \tau\right)
\right]  =\lim_{T\rightarrow\infty}\frac{1}{2T}\left\vert
{\displaystyle\int\limits_{-T}^{T}}
h\left(  t\right)  e^{-i\omega t}dt\right\vert ^{2},
\]
where the Fourier transform is given by%
\[
\mathcal{F}\left[  R_{hh}\left(  \tau\right)  \right]  =%
{\displaystyle\int\limits_{-\infty}^{\infty}}
R_{hh}\left(  \tau\right)  e^{-i\omega\tau}d\tau.
\]
The power spectrum describes how the energy of a signal $h\left(  t\right)  $
is distributed among harmonics in an averaged sense. The total averaged energy
of the signal is given by%
\[
E_{h}=\lim_{T\rightarrow\infty}\frac{1}{2T}%
{\displaystyle\int\limits_{-T}^{T}}
\left\vert h\left(  t\right)  \right\vert ^{2}dt=\overline{\left\vert h\left(
t\right)  \right\vert ^{2}}=R_{hh}\left(  0\right)  =\frac{1}{2\pi}%
{\displaystyle\int\limits_{-\infty}^{\infty}}
S_{hh}\left(  \omega\right)  d\omega.
\]
In contrast to the usual energy spectrum defined by the magnitude of the
Fourier transform of the signal, i.e. $S_{e}\left(  \omega\right)  =\left\vert
\mathcal{F}\left[  h\left(  t\right)  \right]  \right\vert ^{2},$ the power
spectrum can be defined for a signal for which the energy $%
{\displaystyle\int\limits_{-T}^{T}}
\left\vert h\left(  t\right)  \right\vert ^{2}dt$ is not finite. Therefore we
should see the power spectrum as a time or ensemble average of the energy
distributed over different harmonics.

\bibliographystyle{ieeetr}
\bibliography{main}

\end{document}